\documentclass{llncs}

\usepackage{latexsym,url}

\newcommand{\binomials}{\mathcal{E}}
\newcommand{\comment}[1]{}
\newcommand{\ignore}[1]{}

\begin{document}

\title{Enumeration Problems Related to\\Ground Horn Theories}

\author{
Nachum Dershowitz\inst{1} \and Mitchell A. Harris\inst{2} \and Guan-Shieng Huang\inst{3}}

\institute{
School of Computer Science\\
Tel Aviv University\\
Ramat Aviv 69978, Israel\\
\email{Nachum.Dershowitz@cs.tau.ac.il}
\vskip 6pt
\and
Harvard Medical School\\
Department of Radiology\\
Boston, MA 02114, USA\\
\email{harris.mitchell@mgh.harvard.edu}
\vskip 6pt
\and
Department of Computer Science and Information Engineering\\
National Chi Nan University\\
Puli, 545 Nantou, Taiwan R.O.C.\\
\email{shieng@ncnu.edu.tw}
}

\maketitle

\begin{abstract}

We investigate the enumeration of varieties of Boolean theories
related to Horn clauses. We describe a number of combinatorial
equivalences among different characterizations and calculate the
number of different theories in $n$ variables for slightly different
characterizations. The method of counting is via counting models using
a satisfiability checker.

\end{abstract}


\section{Canonical Propositional Systems}
Let $V$ be a set of $n$ propositional variables.  A \emph{nontrivial
monomial} over $V$ is the product (conjunction) of some variables in $V$.  We say
that both $1$ and $0$ are the trivial monomials and take the
convention that $1$ is the product of zero variables.  A
\emph{binomial equation} is an equality of monomials.  Let
$\binomials$ be a set of binomial equations over $V$.  If, say,
$V=\{x, y\}$, we would have
\[ 
   \binomials\subseteq \{xy=x, xy=y, xy=1, xy=0, x=y, x=1, x=0, y=1, y=1, 1=0\}. 
\]
We can apply the Gr\"obner basis construction to $\binomials\cup
\{xx=x\}_{x\in V}$ with a given ordering on monomials, and eventually
we will get a unique canonical system for $\binomials$ and the
particular choice of ordering.  How many distinct canonical systems
are there over $n$ variables?  At first glance, it seems that this
question is very difficult.  There are ${2^n(2^n+1)}/{2}$
$\binomials$'s, and some may have the same canonical system.  We have
to retain those that are in canonical form and rule out those that are
not.

So that this will not be the proverbial search for a needle in a
haystack, we take the following approach: the key point is that this
enumeration problem is in fact equivalent to \emph{counting the number
of distinct Horn functions of $n$ variables}.  A SAT solver (or more
precisely, a SAT enumerator) can help to count this number.  The
following lemma reveals the fact that $\binomials$ indeed defines an
equivalent set of Horn clauses.

\begin{lemma} \label{bin_horn}
    Given any set of binomial equations, there is a set of Horn
    clauses that defines the same constraint.
\end{lemma}
\begin{proof}
    Given any monomial $m$, we use $x\in m$ to denote that variable
    $x$ appears in $m$.  We let $\texttt{true}\in 1$ and
    $\texttt{false}\in 0$.  Let $c(m)$ be the conjunction of elements
    in $m$.  Now for any binomial equation $m_1=m_2$, the
    corresponding Horn clauses are $c(m_1)\Rightarrow x$ for all $x\in
    m_2$ and $c(m_2)\Rightarrow y$ for all $y\in m_1$.
\qed
\end{proof}

Conversely, given any set of Horn clauses, the corresponding set of
binomial equations can be found by the following transformation: Given
$x_1\wedge \cdots \wedge x_j \Rightarrow y$, produce $x_1\cdots x_j=y
x_1\cdots x_j$.

Lemma~\ref{bin_horn} says that each set of binomial equations
corresponds to a presentation of Horn clauses.  Furthermore, there is
a one-one correspondence between the canonical systems of binomial
equations and the Boolean functions that satisfy the system.
Lemma~\ref{bin_horn} implies that these functions are in fact
constrained by Horn clauses.  Let us call a Boolean function
\emph{Horn} if it can be expressed as a conjunction of Horn clauses.
As there are many equivalent sets of binomial equations but only one
is canonical, there are many equivalent presentations made of Horn
clauses but only one Horn function.  We are interested in counting
canonical sets of binomial equations, or canonical Horn presentations.
Either way, the canonical set or canonical presentation is the
canonical representation of a Horn function.  Therefore, our
enumeration problem is identical to count the number of distinct Horn
functions over $V$.

Given a Horn function $f$, we collect the vectors that $f$ maps to $1$
and call that collection the \emph{Horn set} associated with $f$.  Let
$\vec{r}, \vec{s}$, and $\vec{u}$ be vectors in $\{0,1\}^n$.  We
say that $\vec{u}$ is the \emph{meet} of $\vec{r}$ and $\vec{s}$
if $\vec{u}$ is obtained by performing the \emph{logical-and}
operation on each individual coordinate of $\vec{r}$ and
$\vec{s}$.  Horn \cite{AH51} originally characterized
the Horn sets as follows:
\begin{lemma} \label{horn_set}
    A set of vectors is a Horn set iff it is closed under the meet
    operation.
\end{lemma}
Hence, whether or not a set of vectors is Horn can be tested by the
meet criterion of Lemma~\ref{horn_set}.  As there is a one-one
correspondence between Horn functions and Horn sets, our problem is
further reduced to the problem of counting Horn sets.

Now we state how to encode our counting problem in SAT.  For each
vector $\mu$ in $\{0,1\}^n$, we associate with it a predicate
$P_{\mu}$ which means $\mu$ is included in the current Horn set.  For
any $\vec{r}$ and $\vec{s}$, we generate a clause 
$P_{\vec{r}} \wedge P_{\vec{s}} \Rightarrow P_{\vec{u}}$, where
$\vec{u}$ is the meet of $\vec{r}$ and $\vec{s}$.  This set of
clauses asserts the closure property of the meet operation.  Then
we can feed the set of clauses into a \#SAT solver that counts the
number of satisfying models.  

Note that some clauses may be redundant,
since $\vec{u}$ may be identical to $\vec{r}$ or $\vec{s}$.
Therefore, we generate fewer than $4^n$ clauses.


Observe that the generated clauses are indeed Horn clauses.
Therefore, a reasonable DPLL-based \#SAT solver cannot fail to find a
model for any branch of the search, and as one can see the number of
models accumulates quickly.


There are four variations for counting the number of canonical systems
over $n$ variables:
\begin{enumerate}
    \item  $H(n)$ without constants $1$ and $0$ in the systems
           (i.e., no $m=1$ and $m'=0$, where $m$ and $m'$ are monomials);
    \item  $H_{0}(n)$ without constant $1$ (i.e., no $m=1$, but may or may not have $m'=0$);
    \item  $H_{1}(n)$ without constant $0$ (i.e., no $m'=0$, but may or may not have $m=1$);
    \item  $H_{01}(n)$ with both $1$ and $0$ (i.e., may or may not have $m=1$ and $m'=0$).
\end{enumerate}

Counting semi-lattices (idempotent commutative semigroups) with $n$
generators is Case~1; Case~3 is idempotent commutative monoids;
counting the number of Horn theories is Case~4.

There are some relations between these algebras:
\[
\begin{array}{rcl} 
H_{0}(n)&=&2 H(n)\\
H_{01}(n)&=&2 H_{1}(n),
\end{array} \]
since $m=0$ is dual to $m=V$ in the canonical systems, where $m$ is
any monomial and $V$ is the product of all variables.  This fact can
also be seen from their SAT encodings as described above.  The
monomial $m'=0$ forbids the selection of the vector $\vec{1}$ (a vector with
all $1$s) in a Horn set, and selecting the vector $\vec{1}$ in a Horn
set forbids the existence of $m'=0$ in the canonical system.  And in
our SAT encoding, the predicate for $\vec{1}$ (i.e., $P_{\vec{1}}$)
does not appear, since it only occurs in a tautology clause, which is
removed.  Therefore, $P_{\vec{1}}$ is a free variable that can be set
to either true or false, and this is just the case for $H_{0}$ and $H_{01}$.

Case~3 can also be reduced to Case~1, and Case~4 can be reduced to
Case~2, since an equation $m=1$ can also be resolved.  They have the
following relationships:
\[ 
  H_{1}(n) = \sum_{0\leq k \leq n} {n \choose k} H(k), 
\]
\[ 
  H_{01}(n) = \sum_{0\leq k \leq n} {n \choose k} H_{0}(k), 
\]
where ${n \choose k}$ is the binomial coefficient.  Observe that $m=1$
forces all variables appearing in $m$ to be $1$.  Suppose $k$
variables are forced to be $1$.  Those variables can be removed; hence
we get Case~1 and Case~2 systems respectively with $n-k$ variables.  There are ${n \choose
k}$ ways to select $k$ out of $n$ variables to set $1$.  Reversing the direction of summation simplifies the final equation.


We have written a small program that generates clauses that guarantee
closure, which are then sent to a CNF SAT solver to count the total
number of satisfiable truth assignments.  Here are some of the
computed numbers:
\[ 
\begin{array}{r||r|r|r|r|r|r|r|}
 n &~~~~~0& ~~~~~1& ~~~~~2& ~~~~~3& ~~~~~4& ~~~~~5 & ~~~~~6\\\hline\hline
  H(n) & 1& 1& 4& 45& 2271& \mbox{1,373,701} & \mbox{75,965,474,236}\\\hline
  H_{1}(n) & 1& 2& 7& 61& 2480& \mbox{1,385,552} & \mbox{75,973,751,474}\\\hline
\end{array}
\]

\section{Comments}\label{sec:discussion}

Historically, $H_{1}(n)$ counts what is called the number of Moore
families on an $n$-set (Birkhoff \cite{GB67}, citing Moore
\cite{EM10}). A Moore family is a family of subsets that contains the
universal set $\{1,\dots,n\}$ and is closed under intersection. Higuchi
\cite{AH98} computed $H_{1}(n)$ up to $n=5$, directly as Moore
families; Habib and Nourine \cite{HN05} computed the number $H_{1}(6) =
\mbox{75,973,751,474}$ using a correspondence of Moore sets with ideal
color sets of a colored poset.

It has been found that asymptotically $\log_2 a(n) \approx
{n \choose \left\lfloor n/2 \right\rfloor}$ for all of $H(n)$,
$H_{0}(n)$, $H_{1}(n)$, and $H_{01}(n)$ (Alekseev \cite{VA89} and Burosch et
al. \cite{BD93}).

Knuth \cite[Sect.\ 7.1.1]{DK4} (see \cite[\#A108798, \#A108799]{OEIS})
computed the corresponding numbers for the nonisomorphic versions of these systems,
that is, the number of functions distinct under permutation of the
variables. We have not yet discovered a symbolic connection between these enumerations.

\section{Acknowledgements}
Many thanks go to Don Knuth and Neil Sloane for their questions, comments, and references.

\bibliography{enumhorn}
\bibliographystyle{splncs}

\end{document}